\documentclass[10pt, conference]{IEEEtran}

\usepackage[pdftex]{graphicx}
\usepackage{listings}
\usepackage{url}
\usepackage{pgfplots}
\usepackage{pgf-pie}
\pgfplotsset{compat=1.17}
\usepackage{booktabs}
\usepackage{dot2texi}
\usepackage{amssymb}
\usepackage{diagbox}
\usepackage{makecell}

\usepackage[hidelinks]{hyperref}
\usepackage[nameinlink]{cleveref}

\makeatletter
\let\@ORGmakecaption\@makecaption
\long\def\@makecaption#1#2{\@ORGmakecaption{#1}{#2}\vskip\belowcaptionskip\relax}
\makeatother

\usetikzlibrary{shapes, arrows, patterns, automata, positioning}

\newcommand\copyrighttext{%
  \footnotesize \textcopyright 2022 IEEE. Personal use of this material is permitted. Permission from IEEE must be obtained for all other uses, in any current or future media, including reprinting/republishing this material for advertising or promotional purposes, creating new collective works, for resale or redistribution to servers or lists, or reuse of any copyrighted component of this work in other works. Cite this article as follows: L. Sadlek, P. \v{C}eleda, and D. Tovar\v{n}\'ak. \textit{Identification of Attack Paths Using Kill Chain and Attack Graphs}, NOMS 2022-2022 IEEE/IFIP Network Operations and Management Symposium, 2022, pp. 1-6, doi: \href{https://doi.org/10.1109/NOMS54207.2022.9789803}{10.1109/NOMS54207.2022.9789803}.}
\newcommand\copyrightnotice{%
\begin{tikzpicture}[remember picture,overlay]
\node[anchor=south,yshift=12pt] at (current page.south) {\fbox{\parbox{\dimexpr\textwidth-\fboxsep-\fboxrule\relax}{\copyrighttext}}};
\end{tikzpicture}%
}

\begin{document}

\title{Identification of Attack Paths \\Using Kill Chain and Attack Graphs}

\author{\IEEEauthorblockN{Luk\'{a}\v{s} Sadlek\IEEEauthorrefmark{1}\IEEEauthorrefmark{2}, Pavel \v{C}eleda\IEEEauthorrefmark{1}\IEEEauthorrefmark{2}, Daniel Tovar\v{n}\'{a}k\IEEEauthorrefmark{1}}
\IEEEauthorblockA{\IEEEauthorrefmark{2}Faculty of Informatics, Masaryk University, Brno, Czech Republic\\}
\IEEEauthorblockA{\IEEEauthorrefmark{1}Institute of Computer Science, Masaryk University, Brno, Czech Republic\\
sadlek@fi.muni.cz, celeda@ics.muni.cz, tovarnak@ics.muni.cz}
}

\maketitle
\copyrightnotice

\begin{abstract}
The ever-evolving capabilities of cyber attackers force security administrators to focus on the early identification of emerging threats. Targeted cyber attacks usually consist of several phases, from initial reconnaissance of the network environment to final impact on objectives. This paper investigates the identification of multi-step cyber threat scenarios using kill chain and attack graphs.
Kill chain and attack graphs are threat modeling concepts that enable determining weak security defense points.
We propose a novel kill chain attack graph that merges kill chain and attack graphs together. This approach determines possible chains of attacker's actions and their materialization within the protected network. The graph generation uses a categorization of threats according to violated security properties.
The graph allows determining the kill chain phase the administrator should focus on and applicable countermeasures to mitigate possible cyber threats. We implemented the proposed approach for a predefined range of cyber threats, especially vulnerability exploitation and network threats. The approach was validated on a real-world use case. Publicly available implementation contains a proof-of-concept kill chain attack graph generator.

\end{abstract}

\renewcommand\IEEEkeywordsname{Keywords}

\section{Introduction}
The estimated loss from cybercrime in 2020 amounts to 945 billion dollars representing approximately 1\% of global gross domestic product (GDP)~\cite{mcafee2020}. 
Therefore, we focus on timely identification of relevant threats before the attackers exploit defense weaknesses. The identification of cyber threats is related to items of business value for the organization's services called assets, e.g., \textit{information}, \textit{people}, \textit{technologies}, and \textit{facilities}~\cite{certrmm}. The weakness is any bug or flaw in the functionality of used technologies, and vulnerability is its specific instance. Publicly known vulnerabilities, general weakness categories, and types of threats can be obtained from enumerations and knowledge bases. On the other hand, internal information about assets is provided by the organization's asset management.

Multi-step attacks can be modeled using the military kill chain concept. A well-known example is the \textit{cyber kill chain} (also known as \textit{intrusion kill chain}) which consists of ordered phases describing the attacker's progress in achieving actions on objectives~\cite{hutchins2011}. Another threat model are \textit{attack graphs} that depict all possible chains of attacker's actions and their concrete materialization within the organization's network.

Security issues revealed from alerts or vulnerability scanning correspond to only one step from a sequence of attacker's actions. Simple defense mechanisms (e.g., patch management) may not eliminate all network attack paths.
Custom rules are often used for creating these sequences, and the attack paths do not represent possible attacks modeled by the kill chain.

Therefore, our research goal is to identify sequences of adversarial actions that form multi-step threat scenarios which can become attacks. We deal with the following research question: 
\begin{itemize}
    \item[] \textit{Can we merge kill chains and attack graphs to determine targeted cyber threats that jeopardize protected infrastructure and defense against them?}
\end{itemize}

Our contribution can be summarized as follows. We proposed the kill chain attack graph as a new data structure that merges attack graphs and kill chain concepts. We prepared a ruleset for creating attack graphs based on a standardized knowledge base of attack techniques. We proposed a method for chaining these attack techniques into attack paths using violated security properties on assets. We implemented the approach and validated it on a real-world phishing use case.

The paper is organized as follows. Section \ref{sec:rel_work} discusses the current state of the art related to cyber threat identification. Section \ref{sec:kcag} defines the new kill chain attack graph. Section \ref{sec:implement} describes the implementation of attack graph generation. Section \ref{sec:evaluation} provides a validation of the approach. Section \ref{sec:conclusion} concludes the paper.

\section{Related Work} \label{sec:rel_work}
The related work consists of two parts focused on cyber threat identification. These parts are related to the kill chain and the attack graphs.

\subsection{Kill Chains}
The cyber kill chain models attacks as sequences of steps. It assumes that the attacker selects suitable targets first, prepares necessary deliverables, and transmits them to the environment. Then, the deliverables enable to exploit system or application vulnerabilities, install backdoor or trojan, establish a command-and-control channel, and finally act on objectives~\cite{yadav2015}. 

However, it was often criticized for its limited usability, e.g., focus on malware and insufficient support for phishing attacks~\cite{pols}. Therefore, Paul Pols introduced the unified kill chain
consisting of 18 tactics, i.e., phases~\cite{pols}. The tactics are related to establishing a foothold, network propagation, and actions on objectives. The concept uses confidentiality, integrity, and availability as compromised security properties.

MITRE ATT\&CK~\cite{attack} is a knowledge base for attack modeling based on kill chain. ATT\&CK is a matrix consisting of columns and rows. Its tactics (columns) can be viewed as phases of a multi-step attack and techniques (rows) as particular attack steps. Some techniques may not appear in a single multi-step attack and the same order. A recent effort called MITRE D3FEND specifies countermeasures~\cite{d3fend} that can be applied for some ATT\&CK techniques.

\subsection{Attack Graphs}
Attack graphs were introduced by Phillips and Swiler in 1998~\cite{phillips1998graph}. These graphs depict the attacker's paths through the network. The attacker usually accomplishes a sequence of attack steps where each attack step leads to some privileges on protected assets. The popularity of attack graphs manifests in their joint application with other cybersecurity concepts, e.g., Bayesian networks \cite{Poolsappasit2012}.

Research papers usually do not have any standardized attack steps. Exceptions are papers that utilize some well-known enumerations and knowledge bases. A threat modeling language that utilized ATT\&CK was proposed in~\cite{xiong2021cyber}. ATT\&CK and CVE (Common Vulnerabilities and Exposures \cite{cve}) were utilized in~\cite{inokuchi2019design}. Aksu et al. used CVE vulnerabilities from the NVD (National Vulnerability Database)~\cite{aksu2018automated}. ATT\&CK is usually applied as a detailed taxonomy of the attacker's actions and not as a kill chain. To the best of our knowledge, attack graphs cannot represent sequences of attacker’s actions mapped to kill chain phases \cite{kaynar2016taxonomy}.

Generators of attack graphs were already designed and implemented. A well-known attack graph generator is called MulVAL. It takes as input data advisories, host configuration, network configuration, information about users, the interaction of components, and a policy~\cite{ou2005mulval}. However, Kaynar pointed out that the generators can create unrealistic attack paths, e.g., by considering reachability only among hosts and simplifying relationships among applications~\cite{kaynar2016taxonomy}.

\section{Kill Chain Attack Graph} \label{sec:kcag}
Herein, we introduce a new \textit{Kill Chain Attack Graph (KCAG)} and a description of adjustments for the attack graph generation process. \textit{KCAG} is a specific type of attack graph that combines the kill chain and the attack graph concepts. It allows representing chains of attacker's actions divided into kill chain phases.

\subsection{Definition of the KCAG}
\textit{KCAG} is an ordered triple $(G, P, f)$ where $G=(V,E)$ denotes a directed graph with vertices $V$ and edges $E$. A set $P$ contains kill chain phases, and a function $f$ assigns kill chain phases to attack techniques.

Vertices are of five types:
\begin{enumerate}
    \item attacker's level of control over an asset,
    \item property of an asset,
    \item countermeasure,
    \item attack technique,
    \item attack goal.
\end{enumerate}

The first type of vertices expresses the attacker's level of control over an asset. We focus on four categories of assets -- processes, people, technologies, and data. The process is a sequence of actions accomplished by \emph{actors} who utilize technologies to work with data. The process is not accomplished when the security requirements of its related assets are violated. 

Each person owns a collection of accounts and \emph{technologies}, has allowed \emph{actions}, and works with data. Technologies are secondary assets that support processes, e.g., software installed on a network host. Further, we deal with \emph{data at rest} and \emph{data in transit}.

The attacker can have three levels of control over assets. Level zero corresponds to the state when the attacker does not know about the asset's existence. At the first level, the attacker knows about the asset's existence but does not have any rights or cannot violate asset's security properties. At the highest level, the attacker can violate the asset's security requirements. 

The asset type determines potential security properties. Data requires confidentiality, integrity, and availability. Actors are related to authentication and non-repudiation. On the other hand, actions and secondary assets require all security properties from the STRIDE threat model (Spoofing, Tampering, Repudiation, Information disclosure, Denial of service, and Elevation of privilege). These properties are authentication, integrity, non-repudiation, confidentiality, availability, and authorization.

Asset type called \textit{external actor} is always a leaf representing the attacker’s default position (level number zero). Final attack paths contain levels of the attacker's control. Therefore, a list of assets that the attacker accessed or compromised during an attack can be obtained from the attack path vertices. The attacker can also gain control related to the second level from level zero, i.e., skip the first level.

\textit{Properties of assets} are the second type of vertices. They contain information about network services, vulnerable applications, and user accounts. \textit{Countermeasures} are anti-prerequisites for attack techniques. For example, employing a strong password policy hinders the brute force technique. Properties of assets and countermeasures are always leaves.

\begin{table*}[h]
    \centering
    \caption{Selected ATT\&CK techniques mapped to tactics and violated security properties}
    \begin{tabular}{l|l|l|l}
        \toprule
        \textbf{Tactic (Kill Chain Phase)} & \textbf{ATT\&CK ID} & \textbf{Technique Name} & \textbf{Violated Property} \\ \midrule
        Initial Access & T1190 & Exploit Public-Facing Application & Authorization \\
        Initial Access & T1133 & External Remote Services & Authentication\\
        Initial Access & T1566.002 & Spearphishing Link & Authentication\\
        Initial Access, Privilege Escalation & T1078.001 & Default accounts & Authentication\\
        Execution & T1059.008 & Network Device Command Line Interpreters & Authorization\\
        Execution & T1204.001 & User Execution - Malicious link & Authentication\\
        Execution & T1203 & Exploitation for Client Execution & Authorization\\
        Privilege Escalation & T1068 & Exploitation for Privilege Escalation & Authorization\\
        Credential Access & T1110 & Brute Force & Authentication\\
        Discovery & T1046 & Network Service Scanning & Authentication\\
        Discovery & T1018 & Remote System Discovery & Authentication\\ 
        Lateral Movement & T1021 & Remote Services & Authentication\\
        Lateral Movement & T1210 & Exploitation of Remote Services & Authorization\\
        Collection & T1005 & Data from Local System & Confidentiality\\
        Impact & T1499.004 & Endpoint DoS -- Application or System Exploitation & Availability\\
        Impact & T1498 & Network Denial of Service & Availability\\
        Impact & T1489 & Service Stop & Availability\\
        Impact & T1486 & Data Encrypted for Impact & Integrity, Availability\\
        Impact & T1565.001 & Data Manipulation - Stored Data Manipulation & Integrity, Availability\\
        Impact & T1485 & Data Destruction & Integrity, Availability\\ \bottomrule
    \end{tabular}
    \label{tab:techniques}
\end{table*}

\textit{Attack techniques} are vertices with incoming edges from prerequisites -- previous asset control levels, asset properties, and not applied countermeasures. Each attack technique violates some security properties imposed on the assets (see Table \ref{tab:techniques}). The result of the applied attack technique is a different level of control over another asset, a higher control level over the same asset, or a final attack goal. Only some combinations of input and output asset types are allowed. These combinations are depicted in Table \ref{tab:asset_types}. Allowed combinations are denoted by checkmarks and the forbidden by dashes.

\begin{table}[b]
    \centering
    \caption{Allowed input and output asset types for attack techniques}
    \begin{tabular}{l|c|c|c|c|c}
    \toprule
    \theadfont\diagbox[width=7em]{Input}{Output}& \thead{External\\ Actor} & \thead{Actor} & \thead{Secondary\\ Asset} & \thead{Action} & \thead{Data} \\ \midrule
    Ext. Actor & $-$ & $-$ & $-$ & \checkmark & $-$ \\
    Actor & $-$ & $-$ & \checkmark & \checkmark & $-$\\
    Sec. Asset & $-$ & \checkmark & \checkmark & \checkmark & \checkmark\\
    Action & $-$ & \checkmark & \checkmark & \checkmark & \checkmark\\
    Data & $-$ & $-$ & \checkmark & \checkmark & $-$\\ \bottomrule
    \end{tabular}
    \label{tab:asset_types}
\end{table}

An external actor can only take an action that allows him to interact with the protected network infrastructure. No attack step leads to the default position again. The attacker cannot gain control over an actor from another actor without processing some data via actions or by secondary assets. Besides, we do not allow the attacker to gain control over data immediately after gaining control over another data. The rules for asset types were inspired by the data flow diagram used for the STRIDE threat modeling \cite{hernan2006threat}.

Each technique is mapped to the kill chain phase. For this purpose, we use set $P$ and function $f$ from the \textit{KCAG} definition. $P=\{1, ..., n\}$ is a set of kill chain phases $1, ..., n$, which are called tactics in MITRE ATT\&CK. Function $f$ maps set of attack techniques $T$ to their kill chain phases, i.e., $f: T \rightarrow 2^P - \emptyset$. It holds that each technique can be mapped to one or more phases. 

\textit{Attack goals} are vertices representing mission-critical assets as the attacker's objectives. These vertices have only incoming and no outgoing edges in the output graph. They can be expressed using violated security properties of assets on a specified level of control over assets.

Figure \ref{fig:simple_attack_graph} contains an attack graph excerpt for the brute force technique (black vertex) belonging to the credential access kill chain phase. Green vertices are levels of asset control, and the blue vertices are asset properties. The technique requires previously violated authentication of a network connection, i.e., scanned network services. In this example, a user account exists on the SSH service accessible on the TCP port 22. The brown vertex represents that a countermeasure (a strong password policy) was not applied. Last, the authentication of the SSH user account is violated as a result.

\begin{figure}[t]
    \begin{minipage}[l]{0.49\textwidth}
    \centering
    \begin{tikzpicture}[
         ->, 
         >=stealth', 
         shorten >=1pt,
         auto,
         node distance=1.80cm, 
         thick,
         scale=0.50,
         every node/.style={scale=0.700},
         font=\sffamily
     ]
    \tikzstyle{ownRectangle}=[shape=rectangle, draw, minimum size=0.8cm]
    \node[state, color=blue] (3) {3};
    \node[state, color=teal] (2) [above of=3] {2};
    \node[state, color=blue] (1) [above of=2] {1};
    \node[state, color=brown] (4) [right of=3] {4};
    \node[ownRectangle, color=black, label={[font=\small,text=black, label distance=0.2cm, align=center]88:Credential \\ Access}] (5) [right of=2] {5};
    \node[state, color=teal] (6) [right of=5] {6};
    
    \path (2) edge node {} (5)
    (3) edge node {} (5)
    (1) edge[bend right=20] node {} (5)
    (4) edge node {} (5)
    (5) edge node {} (6);
    \end{tikzpicture}
    \par\hspace{1pt}
    \end{minipage}

\begin{minipage}[l]{0.5\textwidth}
\footnotesize
    \label{tab:simple_ag_ver}
    \begin{tabular}{c|l}
    \toprule
    \textbf{ID} & \textbf{Description} \\ \midrule
    1 & A user account on SSH service running on the server.\\
	2 & Violated authentication of SSH network connection to the server.\\
	3 & SSH service on the server accessible on TCP port 22. \\
	4 & Organization does not use a strong password policy.\\
	5 & T1110 - Brute Force.\\
    6 & Violated authentication of SSH service user account on the server.\\
	\bottomrule
    \end{tabular}
\end{minipage}
\caption{The brute force technique against SSH network service credentials.}
\label{fig:simple_attack_graph}
\end{figure}

\subsection{Benefits of the Proposed Approach}
To the best of our knowledge, attack graphs cannot represent sequences of attacker's actions mapped to kill chain phases~\cite{kaynar2016taxonomy}. On the other hand, the kill chains do not capture all possible sequences of attack steps, such as attack graphs. Minor fixes to the current state (such as changed input files) will not help. Therefore, the joint application of these concepts is incorporated directly into \textit{KCAG}.

It is complicated to determine the right level of detail for actions in attack graphs. Our KCAG can be realistic enough because the kill chain enforces proper modeling of particular attack steps. In addition, we use a standardized knowledge base of the attacker's actions (MITRE ATT\&CK).

Kill chain models suggest eliminating strategic places in sequences of steps~\cite{pols}. However, they do not provide theoretical apparatus for such a task. Discussion about defense measures for network hardening is usually accomplished after the graph was generated~\cite{kaynar2016taxonomy}. Therefore, countermeasures of attack techniques are considered directly within the KCAG's definition and during its generation. The most strategic countermeasures can remove the largest count of attack paths.

\section{Implementation} \label{sec:implement}
\begin{figure}[b]
\begin{center}
\includegraphics[width=8.8cm]{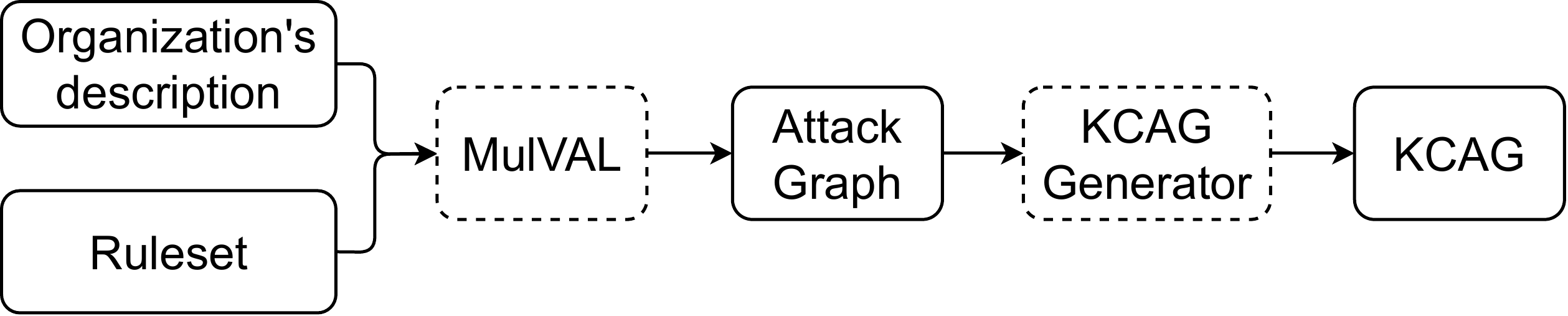}
\end{center}
\caption{Kill chain attack graph generation workflow.} \label{fig:implementation}
\end{figure}

We implemented the proposed approach using the multi-step KCAG generation workflow in Figure \ref{fig:implementation}. One input file describes the organization and the second one contains rules based on ATT\&CK techniques. The attack graph is consequently generated using MulVAL~\cite{ou2005mulval}. Finally, the KCAG generator creates the \textit{KCAG} as a result.

\subsection{Requirements}
The \textit{KCAG}'s design and methodology for creating its vertices must depict the kill chain and the attacker's progress through its phases with their proper ordering. We must automatically determine strategic kill chain phases and countermeasures that can be employed.

We decided to fulfill the first requirement by utilizing a ruleset containing selected ATT\&CK techniques and their kill chain phases.
The ordering of phases is determined by merging the rules into attack paths. Strategic techniques and phases appear in a maximum count of attack paths. These techniques should be addressed by defense measures. 

\subsection{Organization's Description}
The organization's description in a YAML file contains necessary information about secondary assets (i.e., technologies -- hosts, routers, operating systems, software, and network services), network topology, and the organization's missions. Each organization's mission determines requirements on hosts, services, and files. The file also contains a list of identities and user accounts for hosts, services, and domains of hosts.

Vulnerabilities are specified by their CVE IDs. Their impact is categorized into categories specified by our previous work in \cite{komarkova2018community}. These categories are related to code execution, privilege escalation, confidentiality loss, integrity loss, and availability loss. Vulnerabilities are processed differently based on their attack vector from CVSSv3~\cite{cvssv3_1}. Locally exploitable vulnerabilities can be exploited by someone who has access to the system. The attack graph generator considers only vulnerabilities with existing exploits.

Countermeasures specified in the YAML file conform to the D3FEND knowledge graph \cite{d3fend} or were obtained directly from ATT\&CK. Last, the input file contains general information about the organization, e.g., whether employees' emails are publicly available on websites.

Information about technologies, network topology, and user accounts can be obtained automatically, e.g., from asset inventory, using network monitoring, and from application and system logs. Vulnerabilities are obtained from NVD and countermeasures from D3FEND or ATT\&CK. General information about websites can be obtained using web scraping. Last, mission requirements can be obtained manually or automatically depending on the presence of missions in its inventory.

\subsection{Ruleset} \label{subsec:data}
Ruleset can be created in a generic way using the following steps. First, it is necessary to list assets that the defender wants to consider in these categories:
\begin{itemize}
        \item actors (e.g., external actor and user accounts),
        \item actions (e.g., sending an email and network connection),
        \item secondary assets (e.g., operating systems and applications),
        \item data (e.g., file and email message).
    \end{itemize}

Second, we need to prepare syntactic constructs expressing levels of control over these assets. We prepared predicates in the ruleset file. Third, attack techniques should be mapped to kill chain phases and security properties from the STRIDE model as in Table \ref{tab:techniques}. Consequently, countermeasures for the selected techniques (e.g., using D3FEND) should be listed. 

The final step is to prepare rules for attack techniques. The rules should express previous levels of asset control, properties of assets, and possible countermeasures. A result of the attack technique is a level of control over some asset of the type allowed in Table \ref{tab:asset_types}. We manually created a ruleset based on ATT\&CK~\cite{attack}. Its automated creation would require an approach from natural language processing.

Listing \ref{lst:brute_force} contains a rule for the brute force technique. The technique requires a connection to a host \textit{H}. The attacker obtained the second level of its control and violated its authentication by scanning network services. Predicates \textit{networkService} and \textit{hasAccount} identify network service running on the host \textit{H} and the network service's account owner (\textit{Identity}). The strong password policy and multifactor authentication are not used. As a result, the attacker violated the authentication of the network service's user account, i.e., obtained its credentials.

\begin{lstlisting}[language=Prolog, caption={Rule for the brute force technique against network service accounts.}, captionpos=t, belowcaptionskip=4pt, label={lst:brute_force}, basicstyle=\footnotesize, breaklines]
    account(2, authentication, User, Identity, H, Software) :-
        networkConnection(2, authentication, H, Protocol, Port),
        networkService(H, Software, Protocol, Port, _),
        hasAccount(Identity, User, H, Software),
        strongPasswordPolicy(no),
        multifactorAuthentication(no).
\end{lstlisting}

\subsection{Generation of the Attack Graph} \label{subsec:rules}
MulVAL creates the attack graph from the organization's description (i.e., network topology, attack goals, and facts about the organization) and the ruleset. The rules describe how the attacker can escalate control over one asset or move laterally. The generator tries to build an attack path from goals to the initial vertex (in reversed direction) by using rules. The external actor is always the beginning of the attack path.

The ruleset is processed by MulVAL, which determines concrete variable assignment. For example, MulVAL identifies that the brute force rule in Listing \ref{lst:brute_force} is applicable for SSH network service (see Figure \ref{fig:simple_attack_graph}). MulVAL creates the attack graph but does not determine \textit{KCAG}'s vertices types and kill chain phases. We also do not deal with the probabilities of edges. The approach is as scalable as MulVAL's attack graph generation for generating large attack graphs~\cite{ou2005mulval}.

\subsection{KCAG Generator}
The KCAG generator post-processes the attack graph in Python and outputs the \textit{KCAG} as a result. The KCAG generator assigns to each vertex its label according to its type (e.g., countermeasure) and kill chain phases for attack techniques using the mapping function and the set of phases (see Section \ref{sec:kcag}).

The mapping function contains allowed kill chain phases for each technique. Attack paths contain alternating levels of asset control and attack techniques. Therefore, we process subsequent pairs of these attack techniques and their possible phases. We check that assigned phases adhere to their partial ordering. For example, \textit{privilege escalation} cannot be followed by \textit{initial access}. Therefore, if an attack technique belongs to several phases (e.g., \textit{Default Accounts} in Table \ref{tab:techniques}), the KCAG generator allows only phases that adhere to the ordering.

The implementation outputs the attacker's strategic techniques and phases belonging to the maximum count of attack paths. The countermeasures recommended by the KCAG generator can be utilized to destroy all attack paths in the created \textit{KCAG}.

\section{Validation} \label{sec:evaluation}
In this section, we validate that \textit{KCAG} can be utilized in practice. The section describes a real-world multi-step attack containing phishing with malicious attachment and exploitation of two vulnerabilities. Supplementary materials contain all files necessary to reproduce the validation use case \cite{supplementary}.

\subsection{Use Case Setup}
Our setup consists of the organization's description and the ruleset files. The organization's description is available in file \textit{organization.yml}. We chose for clarity a simple setup where the organization's network contains a personal computer connected to the Internet. Windows 8.1 and MS Office are installed on the computer. Their vulnerabilities were obtained from the NVD.

The integrity of files on the computer was specified as a mission requirement. The organization publishes emails and employees' names on their public websites. We specified for each considered countermeasure whether the organization employs it. An employee owns an email account and a user account on the PC.

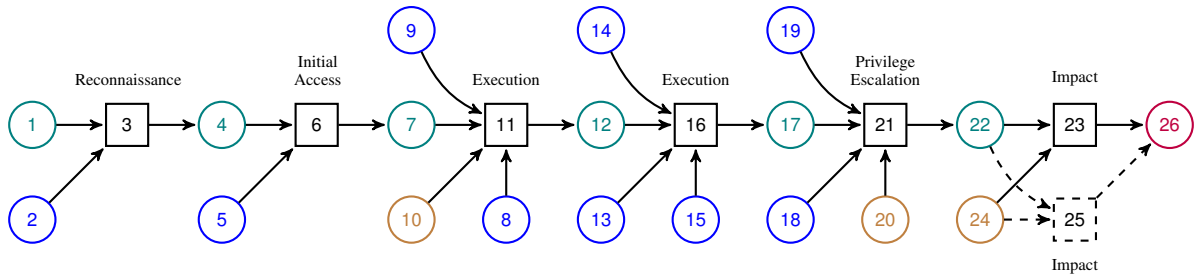
\begin{figure*}[t]
    \centering
    \begin{tikzpicture}[
         ->, 
         >=stealth', 
         shorten >=1pt,
         auto,
         node distance=1.80cm, 
         thick,
         scale=0.50,
         every node/.style={scale=0.700},
         font=\sffamily
     ]
    \tikzstyle{ownRectangle}=[shape=rectangle, draw, minimum size=0.8cm]
    \node[state, color=teal] (1) {1};
    \node[state, color=blue] (2) [below of=1] {2};
    \node[ownRectangle, color=black, label={[font=\small,text=black, label distance=0.2cm, align=center]88:Reconnaissance}] (3) [right of=1] {3};
    \node[state, color=teal] (4) [right of=3] {4};
    \node[state, color=blue] (5) [below of=4] {5};
    \node[ownRectangle, color=black, label={[font=\small,text=black, label distance=0.2cm, align=center]88:Initial \\ Access}] (6) [right of=4] {6};
    \node[state, color=teal] (7) [right of=6] {7};
    \node[state, color=blue] (9) [above of=7] {9};
    \node[state, color=brown] (10) [below of=7] {10};
    \node[state, color=blue] (8) [right of=10] {8};
    \node[ownRectangle, color=black, label={[font=\small,text=black, label distance=0.2cm, align=center]Execution}] (11) [right of=7] {11};
    \node[state, color=teal] (12) [right of=11] {12};
    \node[state, color=blue] (13) [below of=12] {13};
    \node[state, color=blue] (14) [above of=12] {14};
    \node[state, color=blue] (15) [right of=13] {15};
    \node[ownRectangle, color=black, label={[font=\small,text=black, label distance=0.2cm]Execution}] (16) [right of=12] {16};
    \node[state, color=teal] (17) [right of=16] {17};
    \node[state, color=blue] (18) [below of=17] {18};
    \node[state, color=blue] (19) [above of=17] {19};
    \node[state, color=brown] (20) [right of=18] {20};
    \node[ownRectangle, color=black, label={[font=\small,text=black, label distance=0.2cm, align=center]Privilege \\ Escalation}] (21) [right of=17] {21};
    \node[state, color=teal] (22) [right of=21] {22};
    \node[ownRectangle, color=black, label={[font=\small,text=black, label distance=0.2cm]Impact}] (23) [right of=22] {23};
    \node[state, color=brown] (24) [below of=22] {24};
    \node[ownRectangle, color=black, dashed, label={[font=\small,text=black, label distance=0.2cm]270:Impact}] (25) [below of=23] {25};
    \node[state, color=purple] (26) [right of=23] {26};
    
    \path (1) edge node {} (3)
    (2) edge node {} (3)
    (3) edge node {} (4)
    (4) edge node {} (6)
    (5) edge node {} (6)
    (6) edge node {} (7)
    (7) edge node {} (11)
    (8) edge node {} (11)
    (9) edge[bend right=20] node {} (11)
    (10) edge node {} (11)
    (11) edge node {} (12)
    (12) edge node {} (16)
    (13) edge node {} (16)
    (14) edge[bend right=20] node {} (16)
    (15) edge node {} (16)
    (16) edge node {} (17)
    (17) edge node {} (21)
    (18) edge node {} (21)
    (19) edge[bend right=20] node {} (21)
    (20) edge node {} (21)
    (21) edge node {} (22)
    (22) edge node {} (23)
    (22) edge[bend right=20, dashed] node {} (25)
    (23) edge node {} (26)
    (24) edge node {} (23)
    (24) edge[dashed] node {} (25)
    (25) edge[dashed] node {} (26);
    \end{tikzpicture}
    \caption{Example kill chain attack graph for the validation use case. Vertices correspond to attacker's level of control over assets (green), asset properties (blue), countermeasures (brown), attack techniques (black), and an attack goal (dark red). Vertices are described in Table \ref{tab:example_ag_ver}.}
    \label{fig:example_attack_graph}
\end{figure*}

The file \textit{ruleset.P} was created manually. The first part of the ruleset contains predicates that represent instances of \textit{KCAG} vertices except for attack techniques. The attack techniques are specified by rules. MulVAL processes the ruleset and depicts the attack graph in the file \textit{AttackGraph.pdf}.

\subsection{Kill Chain Attack Graph}
The final kill chain attack graph is depicted in Figure \ref{fig:example_attack_graph}, and its vertices are listed in Table \ref{tab:example_ag_ver}. It describes a multi-step attack with the setup mentioned above.
The attacker obtains information about the employees' emails from the organization's website. The attacker sends a convincing phishing email with a malicious MS Word document.

When the curious employee opens the malicious attachment, a code execution vulnerability (\textit{CVE-2017-0262}) is consequently exploited. 
Its exploitation may be accompanied by exploiting the operating system's \textit{CVE-2017-0263}. This CVE allows privilege escalation to administrator privileges. Both vulnerabilities can be exploited locally. These vulnerabilities were exploited in the real world, e.g., by APT28 \cite{attack}.

Finally, the attacker can modify files stored on the personal computer, i.e., encrypt or even destroy them. Modification of files will negatively impact the organization's services. All countermeasures from Figure \ref{fig:example_attack_graph} are suggested as strategic because each one of them can destroy attack paths.

Figure \ref{fig:example_attack_graph} depicts \textit{KCAG} with two attack paths differing in their final attack technique. KCAG generator categorized \textit{KCAG}'s vertices. Moreover, it assigned allowed kill chain phases to attack techniques. The KCAG generator revealed how chains of possible attacker's actions might materialize on protected assets, i.e., used kill chain phases and their ordering.

\begin{table}[t]
    \centering
    \caption{Attack graph vertices}
    \label{tab:example_ag_ver}
    \begin{tabular}{c|l}
    \toprule
    \textbf{ID} & \textbf{Description} \\ \midrule
	1 & External actor.\\
	2 & Email address of an employee was published on a website.\\
	3 & T1594 -- Search Victim-Owned Websites. \\
	4 & The attacker knows that the email address exists.\\
	5 & Sender reputation analysis was not accomplished.\\
    6 & T1566.001 -- Spearphishing Attachment.\\
	7 & Authentication of sending an email was violated.\\
	8 & The employee can click on the attachment.\\
	9 & The employee has a user account on a personal computer.\\
	10 & Training of users was not accomplished (countermeasure).\\
	11 & T1204.002 -- User execution of malicious file. \\
	12 & Authentication of opening file action was violated.\\
	13 & Microsoft Office opens files. \\
	14 & Microsoft Office is installed on the personal computer.\\
	15 & Microsoft Office 2016 contains CVE-2017-0262 vulnerability.\\
	16 & T1203 -- Exploitation for Client Execution.\\
	17 & The attacker violated the system's authorization (user rights).\\
	18 & Microsoft Windows 8.1 is installed on the personal computer.\\
	19 & Microsoft Windows 8.1 contains CVE-2017-0263 vulnerability.\\
	20 & Software is not regularly updated.\\
	21 & T1068 -- Exploitation for Privilege Escalation.\\
	22 & The attacker violated the system's authorization (admin rights).\\
	23 & T1485 -- Data Destruction.\\
	24 & Data backup was not accomplished (countermeasure).\\
	25 & T1486 -- Data Encrypted for Impact.\\
	26 & Integrity of a sensitive file was violated.\\
	\bottomrule
    \end{tabular}
\end{table}

\section{Conclusion} \label{sec:conclusion}
Our research effort focused on cyber threat identification based on asset management data and data about threats.
We proposed a new kill chain attack graph for modeling sequences of the attacker's actions. The kill chain concept only shows the attack lifecycle, and attack graphs do not use any standardized set of actions. Therefore, our adjusted methodology incorporated ATT\&CK techniques and security properties from the STRIDE threat model.

Supplementary materials contain the kill chain attack graph generator in Python \cite{supplementary}. The \textit{KCAG}'s implementation was validated on a real-world use case and a set of techniques. The approach gives the security administrators a way to describe the organization, specify attack rules, and reveal the materialization of multi-step network attacks, including strategic attack techniques and possible countermeasures.

Our future work is to implement generation in an imperative language. It would allow defining asset hierarchy and complicated rules for the attack graph generation. We would also like to use strategic attack steps observed by detection systems to assess the severity of possible multi-step cyber attacks.

\section*{Acknowledgment}
This research was supported by the CONCORDIA project that has received funding from the European Union’s Horizon 2020 research and innovation programme under the grant agreement No 830927.

\bibliographystyle{IEEEtran}
\bibliography{bibliography}

\end{document}